\documentclass[runningheads]{llncs}
\usepackage[dvipsnames]{xcolor}

\usepackage{graphicx}
\usepackage{amsmath,amsfonts,amsbsy}
\usepackage{hyperref}
\usepackage{xcolor}
\usepackage{orcidlink}
\usepackage{mathtools}
\usepackage{bbold}
\usepackage{subcaption}
\usepackage{tikz}
\usepackage{arydshln}

\DeclarePairedDelimiter{\ceil}{\lceil}{\rceil}

\newcommand{\framedlabel}[2]{%
  \begin{tikzpicture}[baseline=(current bounding box.center)]
    \node[draw,rectangle,inner sep=3pt] (box) {\begin{subfigure}{0.8\textwidth}#1\end{subfigure}};
    \node[rotate=90,anchor=south] at (box.east) {#2};
  \end{tikzpicture}%
}

\begin{document}
\title{Operator entanglement growth quantifies complexity of cellular automata}

\titlerunning{OE growth quantifies complexity of CA}
% If the paper title is too long for the running head, you can set
% an abbreviated paper title here
%
\author{Wout Merbis\inst{1,2}\orcidlink{0000-0003-3565-0663} \and
Calvin Bakker\inst{2,3}}%\orcidID{1111-2222-3333-4444}}
\authorrunning{W. Merbis and C.C. Bakker}
% First names are abbreviated in the running head.
% If there are more than two authors, 'et al.' is used.
%
\institute{Dutch Institute for Emergent Phenomena (DIEP) \and Institute for Theoretical Physics (ITFA), \\
University of Amsterdam \\ \email{wmerbis@uva.nl} % \url{https://www.d-iep.org/} 
\and
Instituut-Lorentz, Leiden Institute of Physics, Universiteit Leiden \\
\email{bakker@lorentz.leidenuniv.nl}}
\maketitle              % typeset the header of the contribution
\begin{abstract}
Cellular automata (CA) exemplify systems where simple local interaction rules can lead to intricate and complex emergent phenomena at large scales. The various types of dynamical behavior of CA are usually categorized empirically into Wolfram's complexity classes. Here, we propose a quantitative measure, rooted in quantum information theory, to categorize the complexity of classical deterministic cellular automata. Specifically, we construct a Matrix Product Operator (MPO) of the transition matrix on the space of all possible CA configurations. We find that the growth of entropy of the singular value spectrum of the MPO reveals the complexity of the CA and can be used to characterize its dynamical behavior. This measure defines the concept of \textit{operator entanglement entropy} for CA, demonstrating that quantum information measures can be meaningfully applied to classical deterministic systems.

\end{abstract}

\section{Introduction}

Cellular automata (CA) are models of dynamical complex systems where a large number simple components (cells) are subject to locally defined interaction rules \cite{wolfram1983statistical}. Despite the simple nature of the cells and their interaction rules, surprisingly rich dynamical behavior can emerge at larger scales \cite{wolfram1984cellular,wolfram1984universality}. Known and well-studied emergent properties of CA contain complex pattern formations such as fractals \cite{takahashi1990cellular},  localized excitations (called `lifeforms') and self-reproducing structures \cite{langton1984self}, deterministic chaos \cite{cattaneo1999dynamical} and in some cases the emergence of universal Turing machines allowing for universal computation \cite{cook2004universality}. Due to their relatively simple rules at small scales and the resulting complex behavior on large scales, CA are frequently used as a testing bed for studying emergence in complex systems \cite{langton1990computation}.

The simplest CA with emergent properties are the elementary cellular automata (ECA). These are defined in terms of a one-dimensional array of cells with a binary state space, together with a transition rule depending on the state of the three cell neighborhood that surrounds each cell. Wolfram empirically classified the dynamical behavior of the ECA to fall into four distinct classes \cite{wolfram1984cellular}. Class I CA converge quickly to the uniform state; Class II CA converge quickly to a periodic state; Class III CA show chaotic behavior that does not seem to converge to any regularly repeating pattern; and finally, class IV CA show complex behavior that is \textit{at the edge of chaos} \cite{langton1990computation}; in between the chaotic and periodic states. In this class, localized excitations are found which are able to carry information through the CA and interact with each other over a background periodic structure. 
While the Wolfram complexity classes can (and have) been further refined (see \cite{vispoel2022progress} for a recent review), the classification into four globally different types of dynamical behavior has persisted in the literature. Ultimately, these classifications are done empirically by running the CA from many different initial configurations and observing the resulting patterns. A systematic way to deduce and quantify the complexity of a CA directly from first principles is currently lacking.  

In this work, we use the growth in \textit{operator space entanglement entropy} \cite{zanardi2001entanglement,prosen2007operator}, or simply: the operator entanglement (OE), of the ECA rule under time evolution to quantify its complexity. The OE physically represents the (lossless) compressibility of the transition matrix that evolves CA configurations into the future. To compute the OE, we use methods based on tensor networks \cite{schollwock2011density,verstraete2008matrix,orus2014practical}. Tensor networks decompose large dimensional vectors or matrices as a network of smaller tensors, contracted over an internal (bond) dimension. The bond dimension reflects the compressibility of the large dimensional object. Here, we represent the transition matrix as a \textit{Matrix Product Operator} (MPO), which is a tensor network composed out of a one-dimensional array of tensors. The OE is then defined as the Shannon entropy of the distribution formed from the squares of the singular values across the middle bond of the MPO. This gives a measure of the amount of information flowing between the two halves of the CA.

The OE provides a new, time-dependent measure for the CA's complexity which does not depend on initial conditions, nor on the empirical classification of individual trajectories. Instead, it directly relates the complexity to the ability to compress the high-dimensional transition matrix which contains all possible trajectories of length $t$. 
We find that for the simplest rules, this matrix can be compressed to contain only a single relevant term, while for the most chaotic rules the transition matrix cannot be compressed at all. 
We find that the evolution of the OE in time serves as a powerful indicator of the CA's phenomenological behavior and allows us to classify the CA complexity into four distinct global types. We propose a further refinement within these four types based solely on the characteristics of the OE. 

Previous work using tensor networks has mainly focused on quantum cellular automata (QCA) \cite{arrighi2019overview}.  In \cite{alba2019operator}, the OE growth of a quantum mechanical version of rule 54 was found to grow logarithmic in time. Ref \cite{piroli2020quantum} showed the amount of entanglement which can be created by the action of a QCA is limited by an area law. A quantum version of Game of Life was created in \cite{bleh2012quantum}, which shows the emergence of complexity in the quantum domain, and the entanglement entropy growth for GCA was studied in \cite{hillberry2021entangled,ney2022entanglement}. In this work, we provide for the first time a tensor network analysis for the classical and deterministic Wolfram rules.  We show that the operator entanglement can be used to quantify its complexity, demonstrating the usefulness of quantum information inspired measures in the classical domain.

This work is organized as follows: we start with explaining our construction of the MPO that implements the classical ECA transition rules in section \ref{sec:TNrepresentation}. In section \ref{sec:OperatorEE}, we introduce the operator entanglement for ECA and propose a categorizations based on its dynamical behavior. We conclude in section \ref{sec:Conclusion} and give perspective on future work along the lines of the work presented here.

\section{A Matrix Product Operator for elementary cellular automata}
\label{sec:TNrepresentation}

The one-dimensional, discrete state, discrete time cellular automata are defined on an array of cells, where $x_i^{(t)}$ denotes the state of cell $i$ at time $t$. The state of each cell is given by an integer $x_i \in \{0,\ldots ,k-1\}$. The evolution of the CA is determined by iterating the mapping:
\begin{equation}\label{transfunction}
    x_i^{(t)} = f\left[x_{i-r}^{(t-1)}, \ \ldots \  , x_{i}^{(t-1)}, \ \ldots \ , x_{i+r}^{(t-1)} \right]\,,
\end{equation}
where $r$ is the `range' of the CA. In this work, we will primarily be concerned with the elementary cellular automata (ECA), that have $k=2$ and $r=1$. For the ECA, the transition function \eqref{transfunction} of each cell only depends on a neighborhood containing itself and its immediate neighbors. There are then $2^3 = 8$ possible neighborhood states, which results in $2^{8}= 256$ possible transition rules. These transition rules follow a conventional numbering system, where the 8-bit binary representation of the rule number determines the cells future state for each of the 8 neighbourhood configurations.  
An important detail of the time evolution is that all cells are updated simultaneously, which we will call `parallel updates'. % 

The ECA with transition rules given by \eqref{transfunction} are non-linear dynamical systems, where the state of a cell depends on a multi-linear function of its neighborhood. However, any finite CA of length $L$ can also be represented as a bit string $\mathbf{x}^{(t)} = \{x_1^{(t)} x_2^{(t)} \ldots x_L^{(t)}\}$, which is vectorized into a vector of dimension $2^L$. In this case, the dynamics of the CA is a map from bit string to bit string, that can be implemented as a \textit{linear} operation on the vector space representing all possible bit strings (or: of all \textit{state configurations} of the CA). Hence, we can represent any CA update rule as a matrix $\mathcal{\hat{T}}$ that maps the input CA configuration to an output CA configuration. This transition matrix is $2^L \times 2^L$ dimensional, making its explicit computation for large system sizes intractable. 

\begin{figure}[t]
    \centering
    \includegraphics[width=0.72\textwidth]{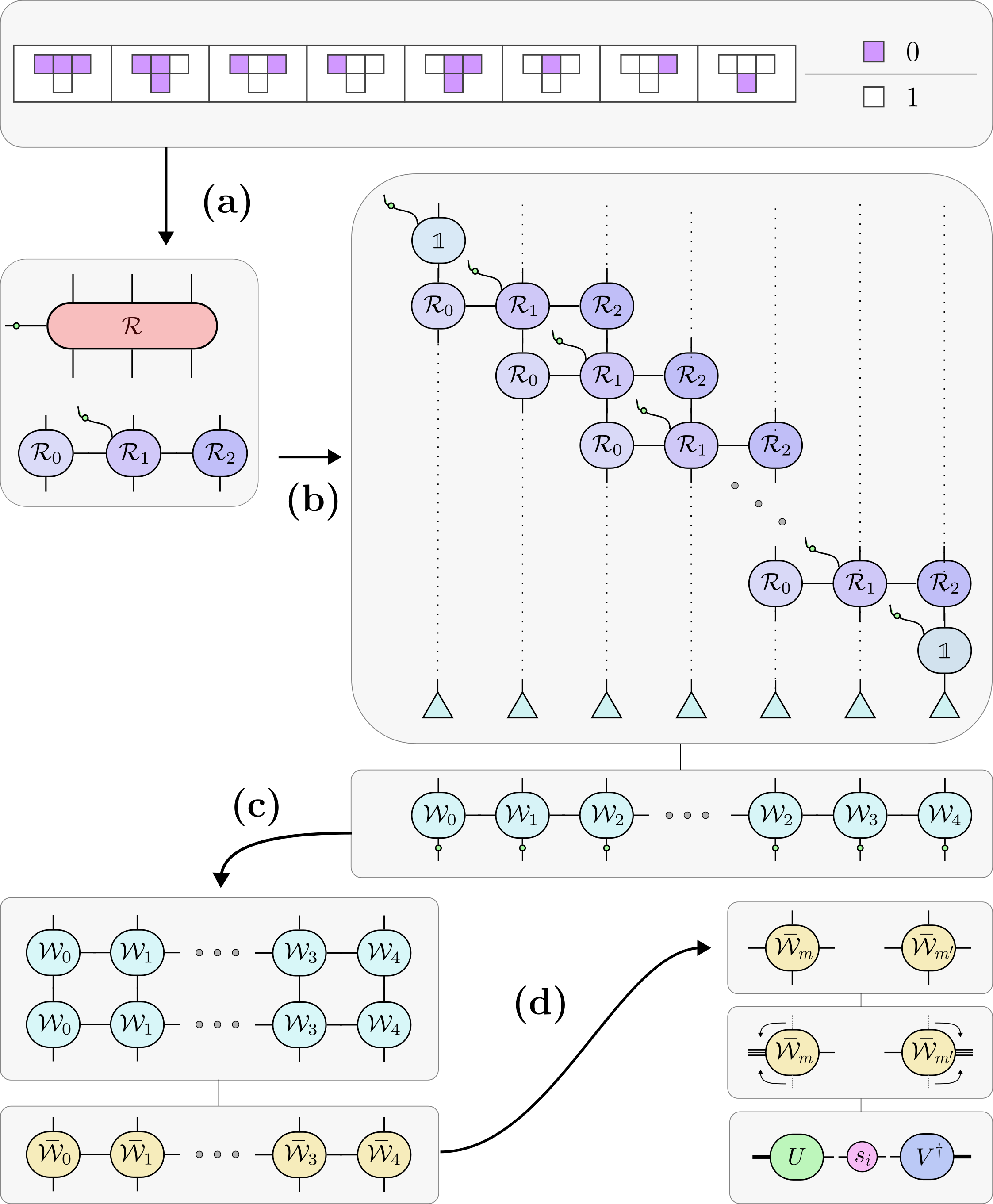}
    \caption{Overview of our methodology: \textbf{(a)} the construction of a tensor $\mathcal{R}$ from the ECA rule, \textbf{(b)} formation of the Matrix Product Operator of the transition matrix $\hat{\mathcal{T}}$, implementing a single time step \textbf{(c)} contraction of MPOs to build $ \hat{\mathcal{T}} \cdot \hat{\mathcal{T}}$, implementing two time steps at once and \textbf{(d)} extraction of singular values $s_i$ from the bonds between the MPO tensors.}
    \label{fig:overview}
\end{figure}

To overcome this `curse of dimensionality' for the single time-step transition matrix, we construct a Matrix Product Operator (MPO) representation for the transition matrix. An MPO can give a lower dimensional (compressed) representation for high dimensional matrices, by decomposing it into an array of smaller rank-4 tensors, that, when contracted over the internal \textit{bond dimension}, will reproduce the high-dimensional matrix. The bond dimension represents physically the amount of compression which can be achieved, as it captures the influence neighboring sites exert on each other. The transition matrix MPO is obtained as outlined in Figure \ref{fig:overview} and described in the steps below. Here tensors are essentially multi-dimensional arrays, and they are indicated in the figure as boxes with their indices shown as edges. Connected (or: contracted) edges indicate inner products over the corresponding array dimensions.

\begin{itemize}
    \item[\textbf{(a)}] From the transition rule \eqref{transfunction}, we construct a rank-7 tensor $\mathcal{R}$ with components: 
    \begin{equation}\label{Rtensor}
        \mathcal{R}^{r_{i-1} r_{i} r_{i+1}}_{o_{i+1} o_{i} o_{i+1}}{}_{w_i} = \delta^{r_{i-1}}_{o_{i-1}}\delta^{r_{i}}_{o_{i}}\delta^{r_{i+1}}_{o_{i+1}} \delta_{w_i f(r_{i-1},r_i,r_{i+1})}
    \end{equation}
    Here, the three indices indicated by ($r_{i-1}, r_i, r_{i+1}$) represent the state of the neighborhood of cell $i$. The three bottom indices $(o_{i-1} o_i o_{i+1})$ are copies of this state, which we need to implement the update rule on the neighboring cells (see (b) below). The index $w_i$ is labeled by the green circle in Fig \ref{fig:overview} and indicates the new state of cell $i$, according to the transition rule \eqref{transfunction}. Using two consecutive \textit{singular value decompositions} (SVDs), the tensor $\mathcal{R}$ is brought into an matrix product form:
    \begin{equation}
        \mathcal{R}^{r_{i-1} r_{i} r_{i+1}}_{o_{i+1} o_{i} o_{i+1}}{}_{w_i} = \sum_{\alpha_1, \alpha_2} (\mathcal{R}_0)^{r_{i-1}}_{o_{i-1} \alpha_1} (\mathcal{R}_1)^{r_i \alpha_1}_{o_i w_i  \alpha_2} (\mathcal{R}_2)^{r_{i+1} \alpha_2 }_{o_{i+1}} \,.
    \end{equation}
    Here the $\alpha_{1,2}$ indices indicate the \textit{bond dimension} of the MPO.
    \item[\textbf{(b)}]  The MPO $\mathcal{\hat{T}}$ is constructed by sequentially updating the sites with $\mathcal{R}$, starting from the left to the right, followed by summing over the output indices $o_1, \ldots o_L$ when all neighbours are updated. This is indicated by the triangles in Fig \ref{fig:overview}.
    In this work, we consider open boundary conditions where the left and right boundary cells are not updated. The final MPO is:
    \begin{equation}\label{mpo}
        \hat{\mathcal{T}} = \sum_{\{\gamma_1 \ldots \gamma_{L-1} \} } (\mathcal{W}_{0})_{o_1 \gamma_1}^{r_1} (\mathcal{W}_1)_{w_2 \gamma_2}^{r_2 \gamma_1 } (\mathcal{W}_2)_{w_3 \gamma_3}^{r_3 \gamma_2 } \ldots (\mathcal{W}_3)_{w_{L-1} \gamma_{L-1}}^{r_{L-1} \gamma_{L-2} }(\mathcal{W}_4)_{w_L}^{r_L \gamma_{L-1} } \,.
    \end{equation}
    Here the $\gamma$ indices are composed out of direct products of the $\alpha$ indices. After the combination of $\alpha$ bond dimensions into the $\gamma$ bonds, another SVD of the $\mathcal{W}$ tensors is performed. At this stage the resulting MPO has a bond dimension that does not exceed $D=4$.\footnote{Here the value of $4$ is a result of the $\log_2 4  = 2$ bits of information which is exerted on each cell by its two direct neighbors.\label{footnote1}}  
    \item[\textbf{(c)}] The MPO $\hat{\mathcal{T}}$ is a compressed representation of the $2^L \times 2^L$ dimensional transition matrix that maps all CA configurations one time step into the future. Time evolution of the CA for several time steps can now be implemented by an MPO $\hat{\mathcal{T}}^t$, where $t$ represents the (integer) number of time steps. This is obtained by contracting the MPO $\mathcal{\hat{T}}$ with itself repeatedly. During this process, the bond dimension of $\hat{\mathcal{T}}^t$ increases.
    \item[\textbf{(d)}] After each contraction with $\mathcal{\hat{T}}$, the MPO representing the time evolution is compressed using SVDs. This way the bond dimension is reduced such that only the relevant linear combinations of configurations contributing to the systems time evolution are taken into account and the operator $\hat{\mathcal{T}}^t$ is optimally compressed without loss of information.
\end{itemize}
We have verified that all ECA rules are implemented correctly using our MPO representation. Furthermore, it is possible to evolve probability distributions over CA configurations in time, without having to perform ensemble averages over simulated trajectories. We have checked the ECA rules are implemented correctly by contracting the MPO \eqref{mpo} with a Matrix Product State representing a uniform distribution over all configurations. The resulting state was matched with the empirical distribution obtained by evolving all possible initial configurations one time step into the future.

\section{Operator space entanglement entropy growth}
\label{sec:OperatorEE}

Now that we have obtained an MPO representation for any ECA transition rule, we can investigate the complexity of the CA under time evolution. We do so by investigating the \textit{compressibility} of the MPO representation of $\hat{\mathcal{T}}^t$ as time increases. At each time step, we perform an SVD over each bond in the MPO, indicated by the $\gamma$ indices in \eqref{mpo}. Since the SVD decomposes the bond into a product of unitary matrices (which implement an orthogonal basis transformation) and the singular values, it reveals which linear combinations of configurations are most relevant in the time evolution of the CA. If all singular values are equally large, the MPO is incompressible and the bond dimension of the MPO is multiplied by 4 at each timestep (see footnote \ref{footnote1}), resulting in exponential growth as $4^t$. In this situation, all configurations are relevant to predict the future state. If only one singular value is non-zero, the CA's evolution can be described as a map into a single configuration and the MPO can compressed into a direct (Kronecker) product of matrices with bond dimension one. Most rules, however, have a singular value spectrum somewhere in between these two extremes. 

To quantify the complexity of ECA, we investigate the Shannon entropy of the distribution formed from the squares of the singular values $s_i$ across the middle bond of the MPO $\hat{\mathcal{T}}^t$:
\begin{equation}\label{svdentropy}
    S_{L/2}(t) = - \sum_{i} s_i^2 \log_2( s_i^2)\,.
\end{equation}
This we call the \textit{operator entanglement} (OE), in analogy with the corresponding measure for operators in quantum many-body systems \cite{zanardi2001entanglement,prosen2007operator}.\footnote{The analogy with quantum systems is more profound, as this measure is exactly the \textit{entanglement entropy} $S_{L/2}(t) = - \mathrm{Tr} \left[\hat{\rho}_{L/2}(t) \log_2 \hat{\rho}_{L/2}(t) \right]$ of a reduced density matrix $\hat{\rho}_{L/2}(t)$, constructed from the partial trace over half the CA cells of the Gram matrix of the time evolution operator: $\hat{\rho}(t) = (\hat{\mathcal{T}}^t)^T \hat{\mathcal{T}}^t$.}   The OE gives a measure of the information transfer between the two halves of the CA as a function of time. The maximal growth of this information transfer is linear, since at each timestep a cell is only influenced by its direct neighbors. This theoretically maximal growth occurs when the singular value spectrum is uniform and the bond dimension grows as $4^t$, such that $S(t) = 2t$. For the simplest CAs, $S(t)$ drops to zero.

We have computed the growth of OE for all 88 distinct ECA rules\footnote{Out of the 256 possible rules, many are related to each other by symmetry (left-right inversion, bit inversion or both), such that there are only 88 unique rules \cite{cattaneo1997transformations}.}. As the bond dimension grows rapidly for some rules, we halt the time evolution when a maximal bond dimension ($D \sim 1000$) is reached. Still, the behavior of the OE for these rules for small system sizes gives a good indicator of the rule's phenomenological behavior. 
In general, we categorize the behavior of the OE into four distinct types, which may have further subdivisions. Within the theoretical lower and upper bounds, we have found the following four types of characteristic OE behaviour:
\begin{itemize}
    \item[I.] \textbf{Constant} and independent of system size, either immediately or after an initial peak due to transients. The peak time does not depend on $L$.  
    \begin{itemize}
    \item[A.] The OE converges to zero. 
    \item[B.] The OE converges to a finite value.
    \item[C.] The OE converges to an oscillating value.
    \end{itemize}
    \item[II.] Growing, followed by a \textbf{peak} at a time $t \sim L$, and then a decrease.  
    \begin{itemize}
        \item[A.] The OE drops to a constant value independent of $L$.
        \item[B.] The OE drops to a constant value dependent on $L$.
    \end{itemize}
    \item[III.] Growing \textbf{linearly} and reaching a \textbf{plateau} that increases with $L$.
    \begin{itemize}
        \item[A.] The OE grows sub-maximally. %linear growth regime is sub-maximal for small $t$
        \item[B.] The OE grows maximally as $S_{L/2}(t) = 2t$. 
    \end{itemize}
    \item[IV.] Growing \textbf{sub-linearly} and reaching a $L$-dependent plateau at late times, without dropping significantly. 
\end{itemize}
We will now discuss each of these possibilities with examples, and show which type of phenomenological behavior of the CA corresponds to which type of behavior of the operator entanglement. The Wolfram rule numbers corresponding to each type of behavior are summarized in table \ref{table1}, where one can also compare with the empirical Wolfram complexity classes.

\begin{table}[t]
\caption{ECA Wolfram rule numbers corresponding to each type of operator entanglement growth.}\label{table1}
\begin{tabular}{|l|p{7cm}|l|}
\hline
Type &  Wolfram rule number & Wolfram classification\\
\hline
I.A. & \colorbox{Peach}{0, 8, 32, 40, 128, 136, 160, 168}, \colorbox{Orchid}{51, and 204} & Class \colorbox{Peach}{I} \& \colorbox{Orchid}{II} \\ \hdashline
I.B. & \colorbox{Orchid}{4, 12, 13, 19, 23, 36, 44, 50, 72, 76, 77}, \newline \colorbox{Orchid}{78, 104, 132, 140, 164, 172, 178, 200, and 232} & Class \colorbox{Orchid}{II} \\  \hdashline
I.C. & \colorbox{Orchid}{1, 5, 28, 29, 33, 108, and 156} & Class \colorbox{Orchid}{II} \\
\hline
II.A. & \colorbox{Orchid}{2, 3, 7, 10, 15, 24, 27, 34, 35, 38, 42, 46, 56}, \newline \colorbox{Orchid}{58, 130, 138, 152, 162, 170, and 184} & Class \colorbox{Orchid}{II} \\  \hdashline
II.B. &  \colorbox{Orchid}{6, 11, 14, 43, 57, 74, 134, 142, 154}, \newline  \colorbox{Salmon}{30}, and \colorbox{SkyBlue}{106} & Class \colorbox{Orchid}{II}, \colorbox{Salmon}{III}  \& \colorbox{SkyBlue}{IV} \\
\hline
III.A. &  \colorbox{Salmon}{18, 22, 45, 60, 122, 126, and 146} & Class \colorbox{Salmon}{III} \\  \hdashline
III.B. & \colorbox{Salmon}{ 90, 105, and 150} & Class \colorbox{Salmon}{III}  \\
\hline
IV. &  \colorbox{Orchid}{9, 25, 26, 37, 41, 62, 73, 94}, \colorbox{SkyBlue}{54, and 110}  & Class \colorbox{Orchid}{II} \& \colorbox{SkyBlue}{IV} \\
\hline
\end{tabular}
\end{table}

\subsection{Type I rules: quick convergence to constant values}

\begin{figure}
  \centering
  \framedlabel{\includegraphics[width=0.9\linewidth]{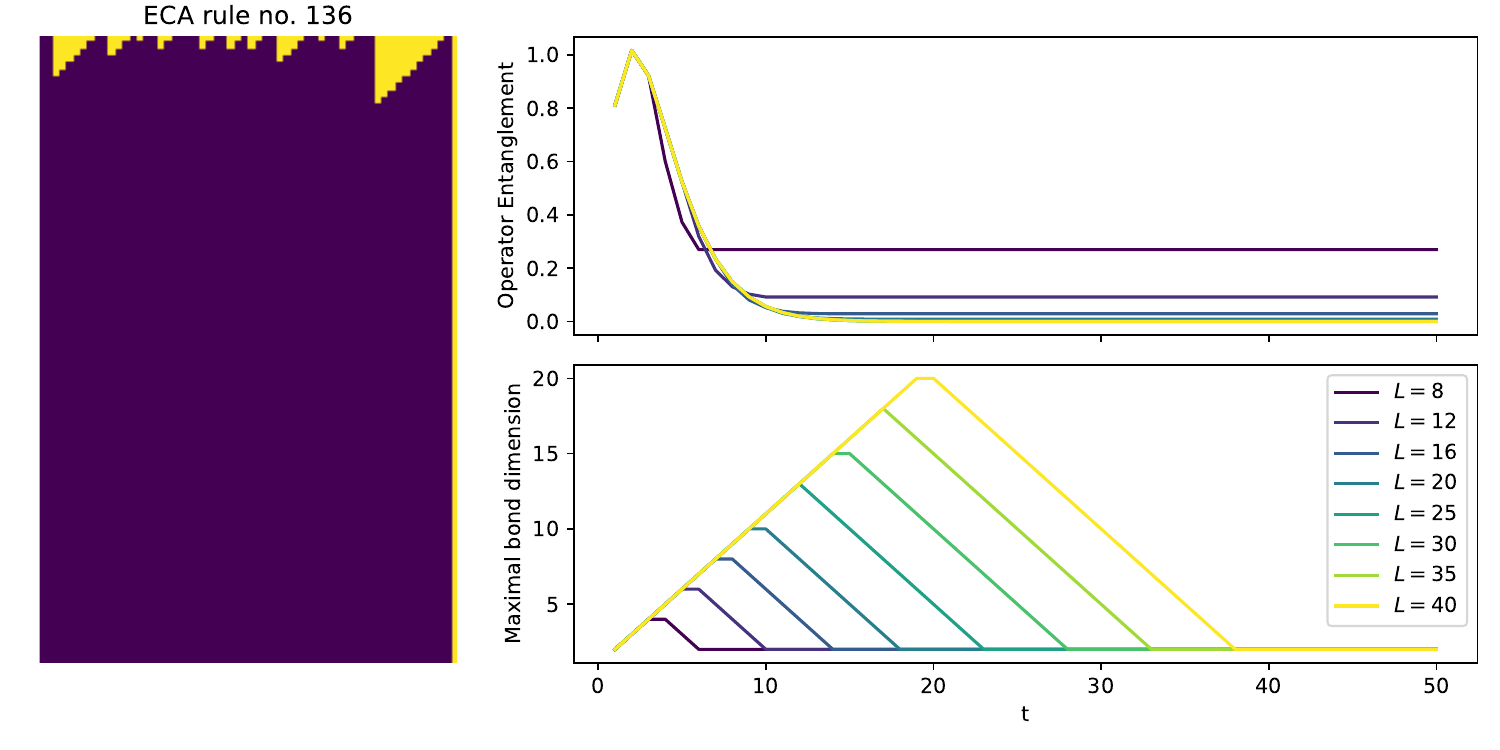}}{Type I.A}
  \framedlabel{\includegraphics[width=0.9\linewidth]{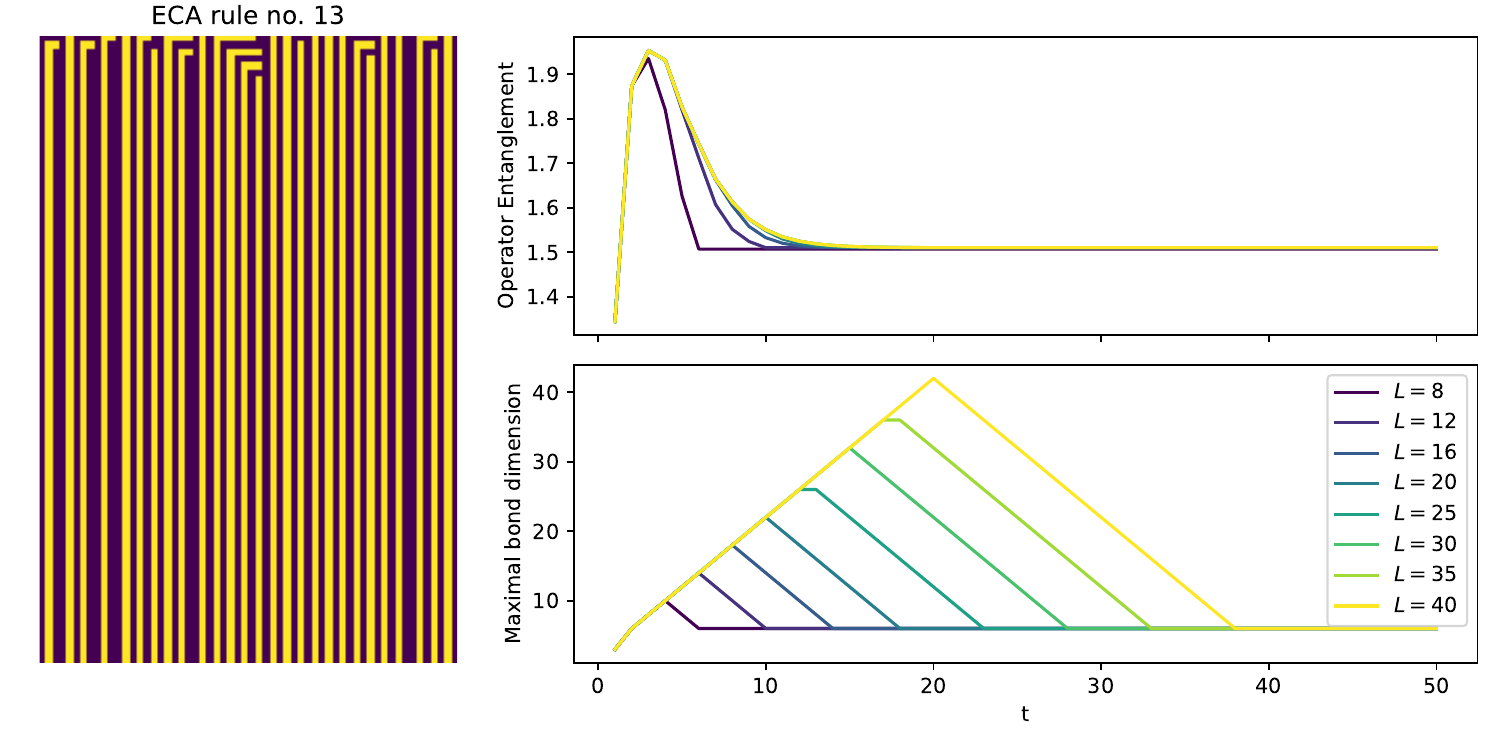}}{Type I.B}
  \framedlabel{\includegraphics[width=0.9\linewidth]{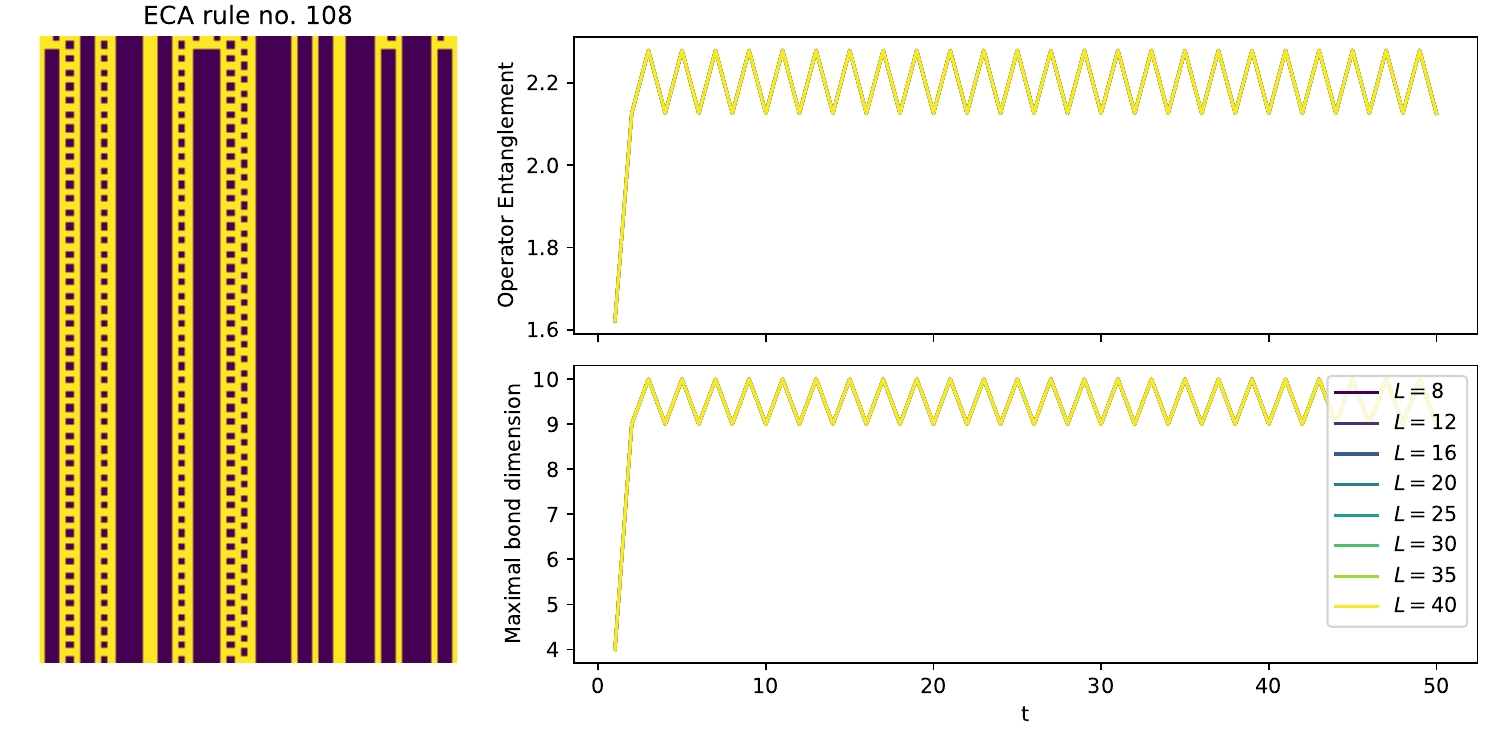}}{Type I.C}
  \caption{Examples for typical OE growth of type I rules of ECA. Each trajectory on  the left is an example of the rules evolution from a random initial configuration. The plots on the right show the operator entanglement and maximal bond dimension of the MPO. These do not depend on the initial configuration and hence say something on the complexity of the dynamical process itself.}
  \label{fig:typeI_examples}
\end{figure}

\subsubsection{Type I.A:}
For type I behavior, the OE quickly converges to a constant. For the simplest rules, this constant is zero. A typical example of this type is shown in the top panel of Figure \ref{fig:typeI_examples} for rule 136. The OE may grow slightly due to transients, but afterwards it quickly drops to zero or a very small value decreasing with $L$. The time of the peak is independent of system size, which signifies that the transients do not lead to long-range information transfer within the CA. 

The ECA rules showing this OE growth all converge to uniform configurations, except for two notable examples: rule 51 and 204. Both rules have vanishing OE, but rule 51 shows oscillatory behavior as it maps each cell to the opposite state regardless of the neighbors state. Rule 204 shows striping patterns, as this rule maps each cells state into itself in the next time step. In both cases, there is no information transfer between neighboring cells. In terms of OE these rules are as complex as rule 0, which maps all configurations to the homogeneous state. 

\subsubsection{Type I.B:}
In the second category of the first type the OE quickly converges to a non-vanishing constant value, as exemplified by rule 13 in the middle panel of Figure \ref{fig:typeI_examples}. By `quickly converging' we mean that if the OE does not immediately reach a constant (and $L$ independent) value, it does so after a peak due to transients that occurs after a time $t_{\rm peak}$ independent of $L$. This implies that the transients do not give rise to long-range information transfers in the CA and the CA quickly settles into either a constant striping pattern, or an oscillating pattern where cells are mapped to its opposing states. In both cases, there are no patterns propagating through the CA, and hence based on the OE these striped patterns and oscillating patterns are equally complex. 

\subsubsection{Type I.C:}
In this category, the OE oscillates between two non-zero values, which do not depend on $L$. The example is illustrated in the bottom panel of Figure \ref{fig:typeI_examples} by rule 108. All rules in this category show the coexistence of striped and oscillation patterns, but there is no long range lateral information transfer of information in the form of propagating shapes through the CA. 

\begin{figure}[t]
  \centering
  \framedlabel{\includegraphics[width=0.9\linewidth]{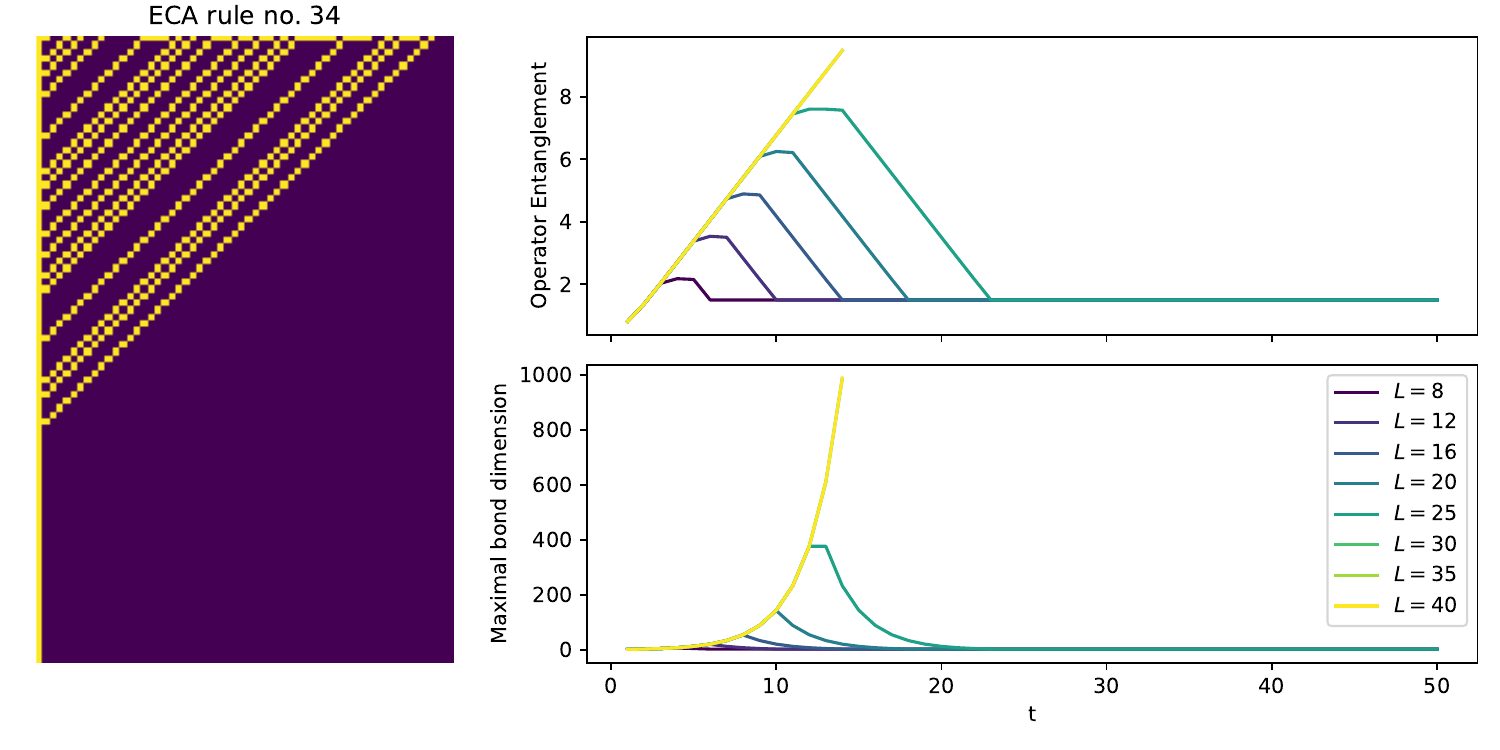}}{Type II.A}
  \framedlabel{\includegraphics[width=0.9\linewidth]{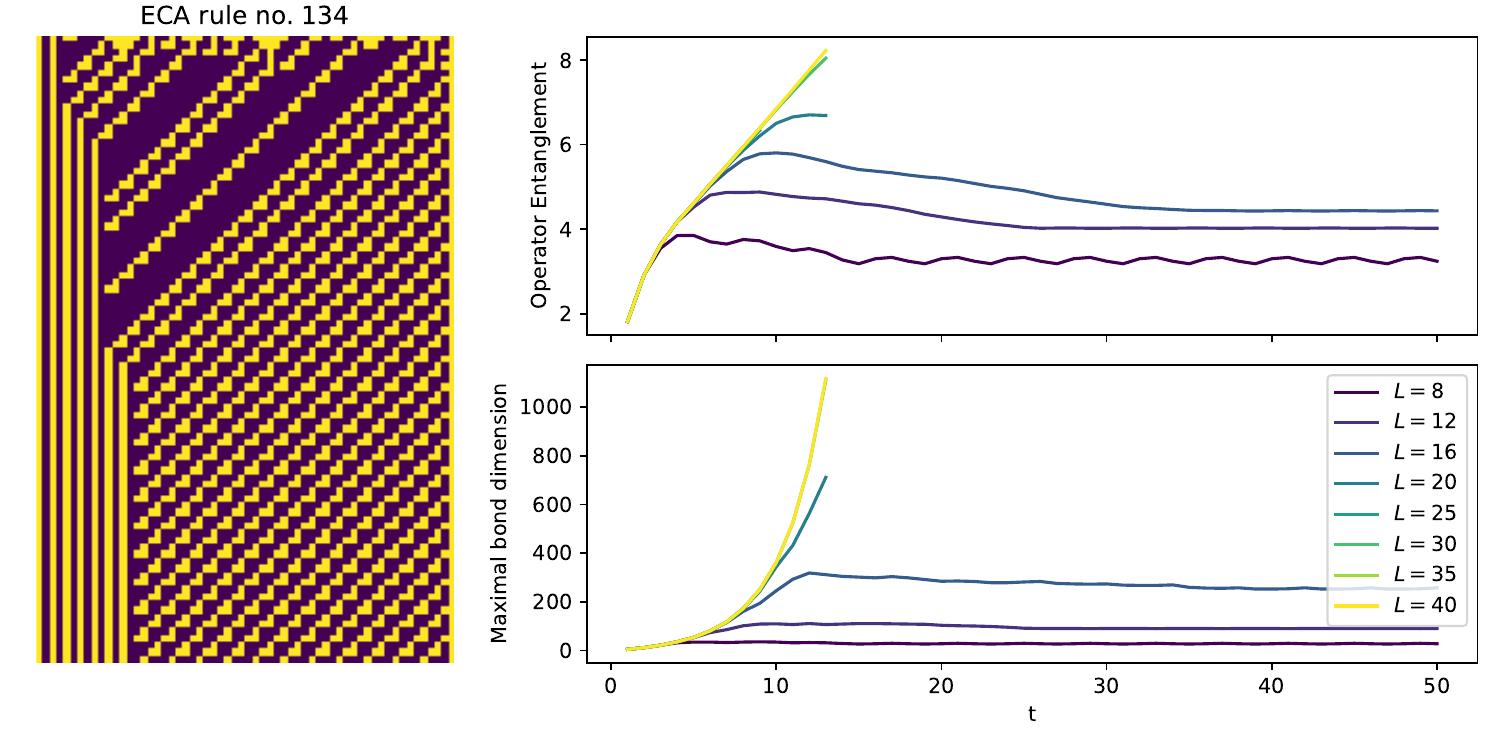}}{Type II.B}
  \caption{Two typical representatives of the second type of OE behavior: rule 34 and rule 134. The computation is performed until a maximum bond dimension is found to be larger than $D \approx 2000$, or numerical convergence issues are encountered.}
  \label{fig:typeII_examples}
\end{figure}

\subsection{Type II rules: long transients and information transfer}
\subsubsection{Type II.A: }

For all rules of the second type, we see the OE initially increase (at most) linearly, however, it reaches a peak after a time $t$ of the order of the system size $L$, after which it decreases again. The distinction we make between type II.A and II.B is whether the OE decreases to a $L$ independent or $L$ dependent constant value, respectively. All rules of type II.A show common phenomenological behavior (see the top panel of Figure \ref{fig:typeII_examples} for an example with rule 34). We observe simple structures propagating linearly through the CA. After a transient time comparable to the CAs size $L$, the system reaches either a homogeneous state, a constant striped state, or an oscillating state with a short period. At this point all propagating structures have reached the boundaries, such that there is no more information being transferred within the CA, leading to a constant OE.

\subsubsection{Type II.B:}

Here the final value of the OE does depend on system size $L$. Phenomenologically, most of these rules are similar to the above type, where initially simple structures propagate through the CA until they reach the boundaries. Only now, the dynamics does not necessarily settle into a simple background pattern, but there may be coexisting domains of different background behaviors (such as stripped and with a part oscillating or a part homogeneous, or a separation between two different oscillating patterns). 

There are also two notable exceptions within this type. Rule 30 is a chaotic rule with periodic boundary conditions, but it shows OE belonging to this type. This is because the open boundary conditions enforce a striping pattern after many time steps. Another exception is rule 106, which has complex phenomenology (class IV), but does show a significant drop in operator entanglement after the initial linear increase. Here, the open boundary conditions enforce a homogeneous state at late times whenever the rightmost cell is empty.

\subsection{Type III rules: chaotic CAs}

\begin{figure}
  \centering
  \framedlabel{\includegraphics[width=0.9\linewidth]{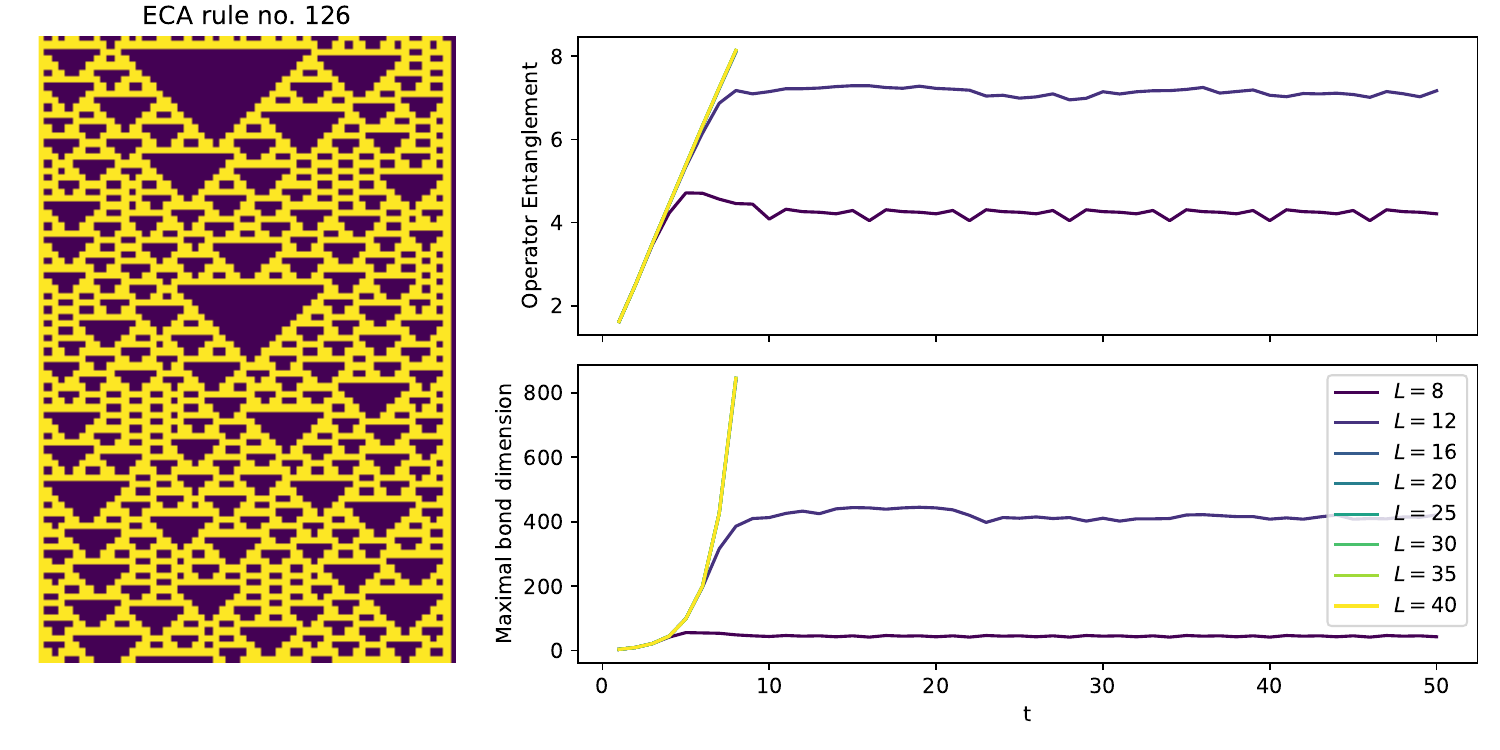}}{Type III.A}
  \framedlabel{\includegraphics[width=0.9\linewidth]{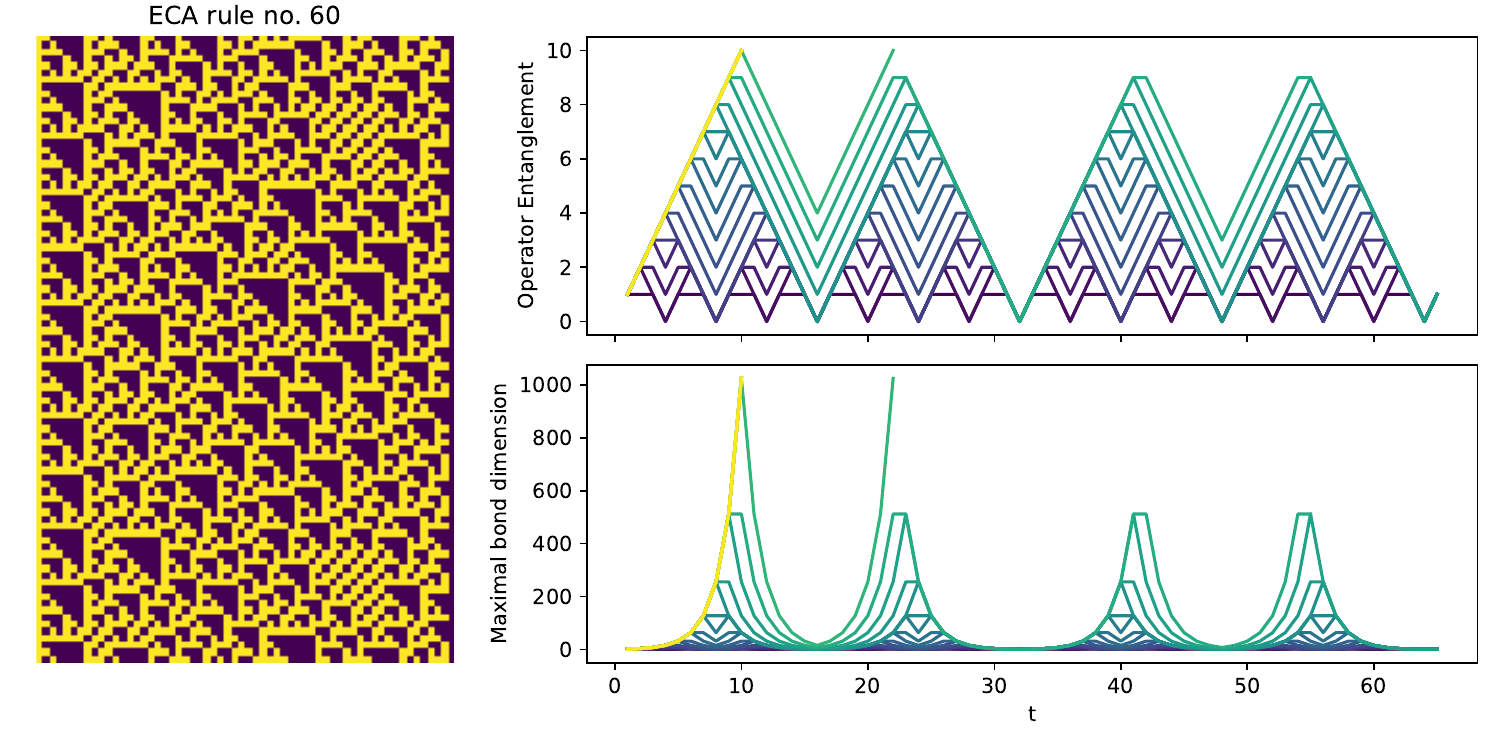}}{Rule 60}
  \framedlabel{\includegraphics[width=0.9\linewidth]{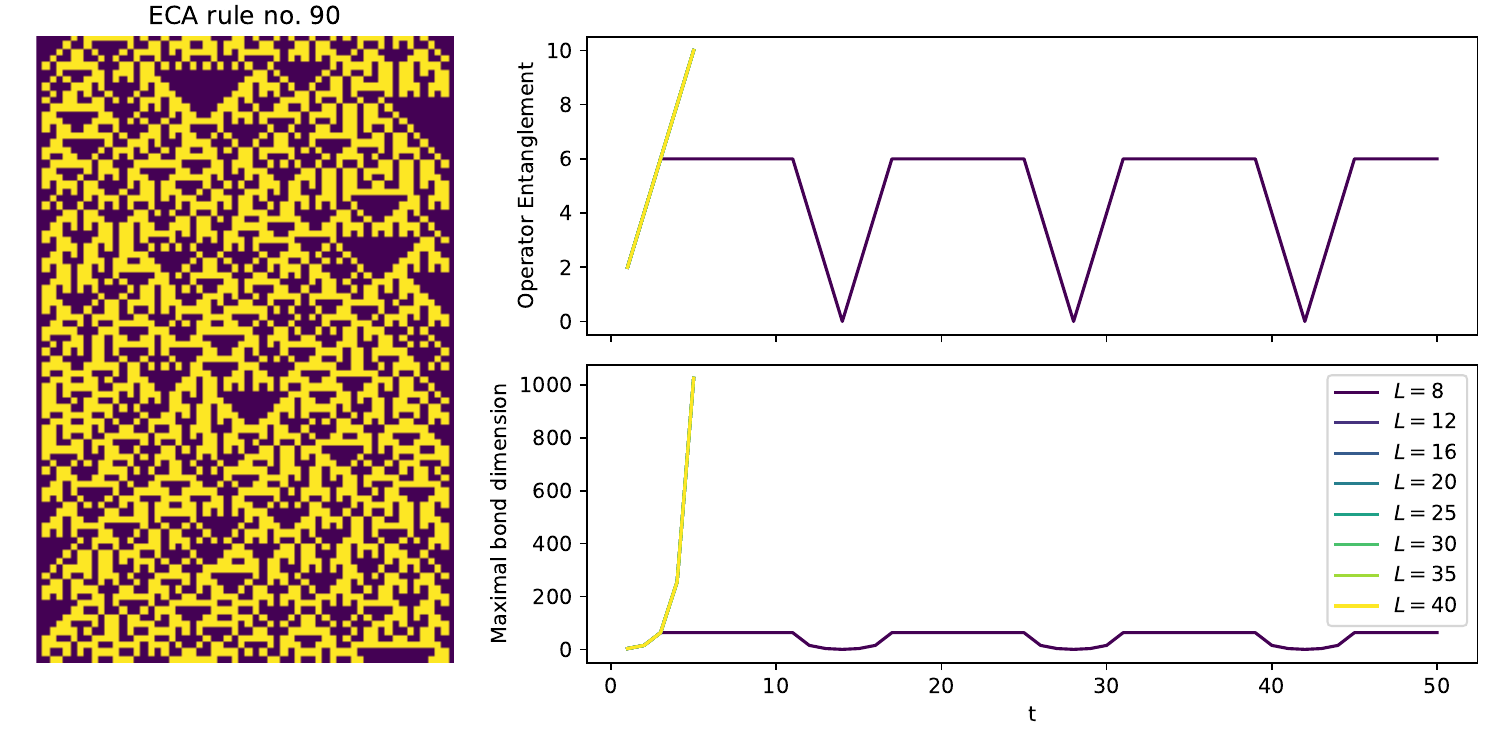}}{Type III.B}
  \framedlabel{\includegraphics[width=0.9\linewidth]{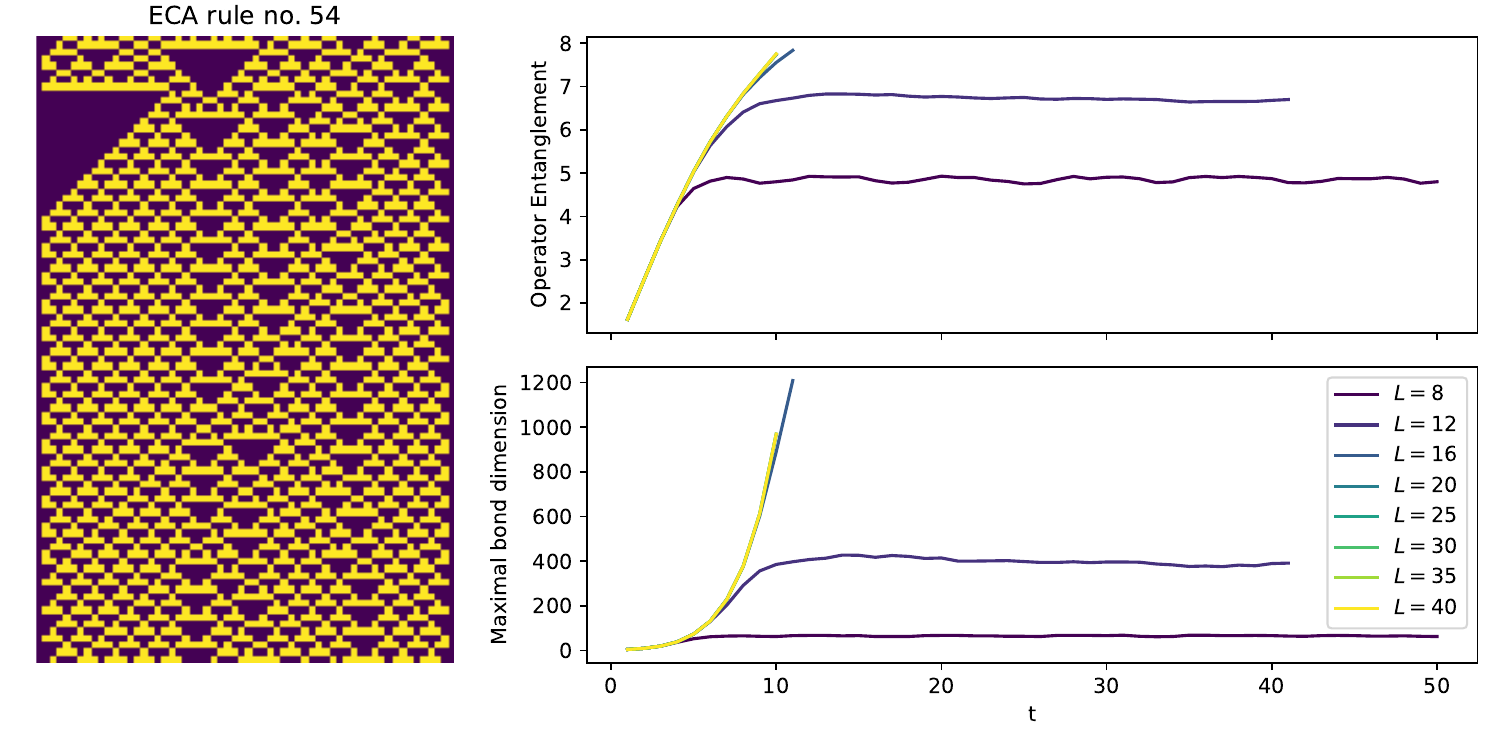}}{Type IV}
  \caption{Representatives of the third and fourth type of OE behavior: rule 126, 60, 90 and 54.}
  \label{fig:typeIII_IV_examples}
\end{figure}

\subsubsection{Type III.A:}

This type of CA produces an exponential increase in bond dimension, and therefore it is computationally hard to propagate these rules over long time scales. For these rules we observe an initial linear growth of OE, which reaches a plateau when $t$ becomes of the order of $L$ and barely decreases afterwards. This type of behavior is exemplified by rule 126 in the top panel of Figure \ref{fig:typeIII_IV_examples}.
All of these rules are characterized by chaotic behavior starting from random initial conditions. This type of behavior is exactly the same as the OE growth in quantum chaotic systems  \cite{prosen2007efficiency,zhou2017operator,dubail2017entanglement}, demonstrating that linear OE growth indicates chaos not only quantum mechanically, but also in classical, deterministic systems. 

An interesting and unique behavior of OE growth is seen for rule 60 (Figure \ref{fig:typeIII_IV_examples}, second panel). Here the initial growth is linear, as a sign of chaotic behavior, but then the OE decreases linearly to a minima at multiples of four time steps. In fact, rule 60 with open boundary conditions is periodic with a period $\tau$ depending on $L$ as $\tau(L) = 2^{\ceil{\log_2(L-1)}}$, \textit{regardless of the initial conditions}. The CA returns to a configuration close to the initial configuration whenever the OE has a minimum, and it returns to the exact starting configuration at times when the OE drops to zero.

\subsubsection{Type III.B:}
For this type the initial growth in OE is linear, but distinctive for these rules is that the growth rate is maximal, such that $S_{L/2} = \log_2(4^t) = 2t$ (see the third panel in Figure \ref{fig:typeIII_IV_examples} for an example). This implies that in these rules, all singular values in the MPO representation are relevant and equally large. In other words, the time evolution operator cannot be compressed and all configurations are relevant for determining the systems future evolution. In these cases, the transition matrix becomes a permutation matrix, where each configuration is mapped one-to-one to another configuration.
Just as is observed for rule 60, for these rules the OE drops to a value of zero periodically.
For instance, for $N=8$, all rules of this type return to the initial configuration after 14 time steps, regardless of initial configuration. This hints at a form of synchronized behavior, even for the apparently most chaotic rules. For finite system sizes, the system does not explore all possible configurations before returning to the initial configuration, but rather gets trapped in cycles of length $t \sim L$. So, rules of this type have a transition matrix composed out of cyclic permutations with uniform cycle length.

\subsection{Type IV rules: domain walls, lifeforms and complexity}
For the final type, we see an operator entanglement that increases sub-linearly (either as $t^{\alpha}$ with $\alpha < 1$ or as $\log(t)$) and then reach a plateau. The plateau height increases with system size. Phenomenologically, these rules are characterized by complex pattern formations. There may be areas of repeating patterns with local excitations (lifeforms) on top of these patterns. The excitations can also function as domain walls between areas with different background patterns, and they may interact in non-trivial ways. The famous rule 110, which is Turing complete, falls into this type. It is worth noticing that many of these rules ultimately settle into a periodically repeating pattern, implying that strictly taken they belong to the Wolfram class 2 cellular automata. The classification we are considering here concerns the initial growth of the operator entanglement, which is sub-linear for all of these rules, and the phenomenology shows there is complex behavior in the transients for all of these rules.     

\subsection{Parallel with quantum systems} 
\begin{figure}[t]
    \centering
    \includegraphics[width=\textwidth]{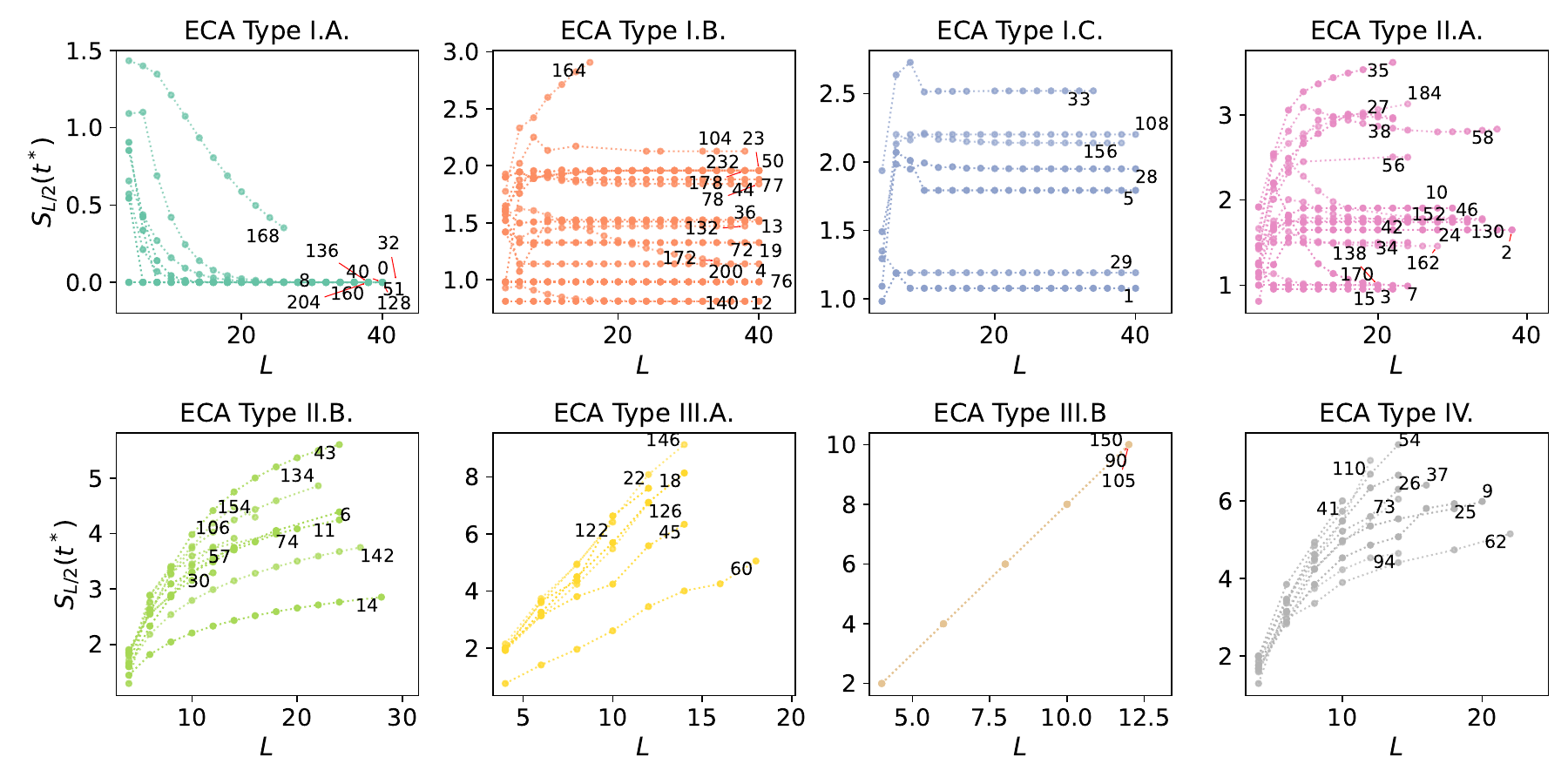}
    \caption{The operator entanglement after converging to a steady value plotted against systems size $L$ for all unique ECA rules. Each subfigure shows the ECA rules of the corresponding type. Curves are labelled by their rule number.}
    \label{fig:OEvsL}
\end{figure}

Figure \ref{fig:OEvsL} shows the value where the OE saturates for all different ECA rules as a function of $L$. There is a clear parallel with the area law scaling of entanglement entropy \cite{schuch2008entropy} in quantum systems. The top panels in Figure \ref{fig:OEvsL} satisfy an `area law',  which in our case implies the OE is (ultimately) constant in system size. For chaotic systems, the OE grows with the volume of the system, in analogy with entanglement growth in quantum chaotic systems \cite{prosen2007efficiency,zhou2017operator,dubail2017entanglement}. The CA rules that show complex behavior of type IV have OE growing sub-linearly in both $t$ and $L$. This is reminiscent of quantum mechanical systems at criticality \cite{dubail2017entanglement} or integrable quantum systems \cite{prosen2007efficiency,alba2019operator}. We wish to explore this parallel in more detail in future work.

\section{Conclusion}
\label{sec:Conclusion}

We have investigated the \textit{operator space entanglement entropy} growth of classical, deterministic cellular automata by mapping the transition rule to a matrix product operator (MPO).
This provides an indicator of the complexity of the ECA rule, as it quantifies information transfer within the ECA, regardless of initial conditions.  
We find that the operator entanglement growth curves can be used to classify ECA with open boundary conditions. We distinguish four main types of behavior of the OE under time evolution, with further refinements for some types, and relate this to the phenomenology of the ECA. We find that the OE either settles to a constant value relatively quickly (type I), or after an initial growth period, displaying a peak at times increasing with system size (type II).  Type III is characterized by linear growth leading to a plateau with height depending on $L$, in analogy with quantum chaotic systems. Type IV rules also have OE growing towards an $L$ dependent plateau, but here the initial growth is sub-linear. 

Our work enables new insight into the information transfer within the CA and thus the inner structure and complexity of a CA rule. It furthermore opens the door to apply the rich toolbox of tensor networks \cite{orus2014practical} to classical CA and other non-linear discrete dynamical system, introducing new computational methods into the study of complex dynamical systems. Possible extensions are the use of density matrix renormalization group algorithms for stochastic (noisy) CA and a study into the large deviation statistics of classical cellular automata \cite{buvca2019exact}.
We wish to explore these topics in future work.

%\bibliographystyle{splncs04}
%\bibliography{biblio}

\end{document}